\documentclass[twocolumn,showpacs,preprintnumbers,amsmath,amssymb]{revtex4}

\usepackage{graphicx}
\usepackage{dcolumn}
\usepackage{bm}
\usepackage{graphicx}
\input{epsf}

\newcommand{\beq}{\begin{equation}}
\newcommand{\eeq}{\end{equation}}
\def\la{\hbox{\raise.35ex\rlap{$<$}\lower.6ex\hbox{$\sim$}\ }}
\def\ga{\hbox{\raise.35ex\rlap{$>$}\lower.6ex\hbox{$\sim$}\ }}

\def\beq{\begin{equation}}
\def\eeq{\end{equation}}
\def\beqa{\begin{eqnarray}}
\def\eeqa{\end{eqnarray}}
\def\bseq{\begin{subequations}}
\def\eseq{\end{subequations}}

\begin{document}

\title{High-growth-rate magnetohydrodynamic instability in differentially rotating compressible flow}

\author{Mradul Sharma}
\affiliation{Theoretical Astrophysics Section, Astrophysical Sciences Division, Bhabha Atomic Research Centre, Mumbai - 400085, India}
\email[]{mradul@barc.gov.in}
\date{\today}

\begin{abstract}
The transport of angular momentum in the outward direction is the fundamental requirement for accretion to proceed in an accretion disk. This objective can 
be achieved if the accretion flow is turbulent. Instabilities are one of the sources for the turbulence. We study a differentially  rotating compressive 
flow in the presence of non vanishing radial and azimuthal magnetic field and demonstrate the occurrence of a high growth rate instability. This instability 
operates in a region where magnetic energy density exceeds the rotational energy density. 
\end{abstract}

\maketitle

\section{Introduction}
Disk accretion is one of the most fundamental processes occurring in a variety of astrophysical objects like accretion powered x-ray pulsars, protostars, 
cataclysmic variables etc. It is observed that the molecular viscosity is too small to account for the observed time scales and luminosities \cite{Pringle}. 
To circumvent this problem, Shakura and Sunyaev \cite{Shakura} made a normalization of the coefficient of viscosity in the theory of turbulent 
viscosity proposed by Heisenberg \cite{Heisenberg} and introduced a dimensionless parameter $\alpha c_s H$, where $c_s$ and $H$ were the sound speed and the 
disk scale height respectively. Though, the $\alpha$ model is quite successful in explaining many astrophysical observations, a degree of ad-hocness enters 
the model through the parameter $\alpha$. The presence of Rayleigh criterion of hydrodynamic stability, i.e.  a flow with specific angular momentum increasing 
monotonically, being satisfied in accretion disk, rules out the hydrodynamic origin of turbulence, though there is literature available \cite{Bani1, Bani2} in its favor. The origin of turbulence is still a question. Differential rotation can lead to an instability provided the specific angular momentum decreases outward, a condition generally not satisfied in the accretion disk, though there are reports suggesting such regions being realized locally \cite{Honma,Manmoto, Kato}. A recent work \cite{Gracia} numerically  demonstrated the presence of such regions. Though differential rotation alone can not lead to instability, inclusion of magnetic field alters the scenario altogether. It was well known from the classical works of Velikhov \cite{Velikhov} and Chandra \cite{Chandra} that a differentially rotating flow with a negative angular velocity gradient in a weak magnetic field is unstable. The presence of this instability in accretion disk was established by Balbus \& Hawley \cite{Balbus} after three decades that the subthermal fields in the presence of differential rotation can cause magnetorotational Instability (MRI), which can provide the physical basis for the transport of angular momentum in accretion disk. It has been computationally shown \cite{Hawley1,Hawley2} that MRI is capable of introducing turbulence in the accretion disk. It is to be noted that MRI is essentially an instability of incompressible flow and its growth starts suppressing as the compressibility 
seeps in. The earliest attempt to understand the effect of compressibility on the growth of MRI was made by  Blaes \& Balbus \cite{Blaes}. They concluded that the $B_{\phi}$ does not affect the instability. Later on, the behavior of compressible MRI in the MHD flows was addressed by Kim \& Ostriker \cite{kim}. It was demonstrated that when the magnetic field strength is superthermal, the inclusion of toroidal fields tends to suppress the growth of the MRI, and that for quasi-toroidal field configurations no axisymmetric MRI takes place in the limit $c_s \rightarrow 0$. In an another work, Pessah \& Psaltis \cite{Pessah} studied the role of toroidal fields in compressible flows. It was again demonstrated that the growth rate of MRI is affected in the compressible flow and MRI stabilizes in superthermal fields.  Clearly, the 
compressibility plays an important role in the dynamics of instabilities operating in the accretion flow.

Recently, Bonanno and Urpin \cite{BnU06} (henceforth paper I) studied the effect of compressibility on the instabilities in the presence of magnetic field. They considered the magnetic field with non vanishing radial and azimuthal components. A new instability was observed which survived for all  values of magnetic field, unlike MRI which survives only in the weak field limit. The maximum growth rate of the reported instability was $\sim \Omega$ where $ \Omega$ is the rotation frequency. In a 
recent work \cite{Mradul}, we investigated the behavior of above instability for a special case $\Omega_e^2<0$ ($\Omega_e^2$ being epicyclic frequency, defined
 as  $ \Omega_e^2  =  4 \Omega^2 (1 + \frac{s}{2 \Omega} \frac{d \Omega}{d s}$), $\Omega$ is the rotational frequency). An instability with a  high growth rate was 
observed. In an offshoot of this work, we investigate the behavior of instability reported in \cite{Mradul} in the Keplerian flow. It is to be noted that the only difference between the work carried by Bonanno and Urpin \cite{BnU06} and us is the inclusion of parameter space which was not considered in their study.

The  paper is organized as follows: The section II investigates the growth rate of instability. The section III deals with the Results and Discussions. Finally, we conclude by summarizing the new findings.

\section{The instability criteria}
Paper I considered  an axisymmetric differentially rotating system in the presence of a magnetic field.  A cylindrical coordinate system ($s$, $\varphi$, $z$) with s being the radial distance from the rotation axis was constructed. Unperturbed system was described by ($v_r, v_{\phi}, v_z $) = ($0, s \Omega, 0$). Furthermore, $\Omega$ where $\Omega$ being the angular velocity of the astrophysical flow, was taken to be approximately a function of s alone; i.e. $\Omega = \Omega (s)$.  The isothermal flow for a compressible fluid was described by the MHD equations

\begin{eqnarray}
\dot{\vec{v}} + (\vec{v} \cdot \nabla) \vec{v} = - \frac{\nabla p}{\rho} 
+ \vec{g} + \frac{1}{4 \pi \rho} (\nabla \times \vec{B}) \times \vec{B}, 
\end{eqnarray}
\begin{equation}
\dot{\rho} + \nabla \cdot (\rho \vec{v}) = 0, 
\end{equation}
\begin{equation}
\dot{p} + \vec{v} \cdot \nabla p + \gamma p \nabla \cdot 
\vec{v} = 0,
\end{equation}
\begin{equation}
\dot{\vec{B}} - \nabla \times (\vec{v} \times \vec{B}) + \eta
\nabla \times (\nabla \times \vec{B}) = 0,
\end{equation}
\begin{equation}
\nabla \cdot \vec{B} = 0. 
\end{equation} 
where $\rho$ and $\vec{v}$ are the density and fluid velocity, respectively; $p$ is the gas pressure; $\vec{g}$ is gravity; $\vec{B}$ is the magnetic field, $\eta$ is the magnetic diffusivity, and $\gamma$ is the adiabatic index. Magnetic field has non vanishing radial and azimuthal components.

Quasistationary state for a differentially rotating flow with magnetic field was considered. Axisymmetric Eulerian perturbations with space time dependence $\propto \exp ( \sigma t - i \vec{k} \cdot \vec{r})$ where $\vec{k}= (k_{s}, 0, k_{z})$ were introduced in the unperturbed accretion disk and the dispersion relation was obtained for a special case of $\vec{k} . \vec{B} = 0$ after neglecting the Ohmic dissipation in the induction equation.

We take the dispersion relation of paper I for determining the stability of axisymmetric short wavelength perturbations . The dispersion relation (Eq. ($16$) of paper I) is given by the  
\begin{equation}\label{e:eq1}
\sigma^{5} + \sigma^{3} (\omega^{2}_{0} + \Omega^{2}_{e})
+ \sigma^{2} \omega^{3}_{B \Omega} + \sigma \mu \Omega^{2}_{e} 
\omega^{2}_{0} + \mu \Omega^{2}_{e} \omega^{3}_{B \Omega} = 0
\end{equation}
where
\begin{eqnarray}
\Omega^{2}_{e} = 2 \Omega
(2 \Omega + s \frac{d \Omega}{ds}) \;, \;\;
\omega^{2}_{0}= k^{2} ( c^{2}_{s} + c^{2}_{m})\; ,\;\;
\mu = k^{2}_{z} / k^{2}\;,
\nonumber\\
c^{2}_{m} = \frac{B^{2}}{4 \pi \rho} \;, \;\; 
c^{2}_{s} = \frac{\gamma p}{\rho} \;, \;\; \omega^{3}_{B \Omega}=
\frac{k^{2} B_{\varphi} B_{s} s \Omega'}{4 \pi \rho}  \ ,\Omega' = \frac{d \Omega}{ds};
\nonumber 
\end{eqnarray}

Equation (\ref{e:eq1}) is a polynomial of degree five, so five non trivial roots exist. We apply Routh-Hurwitz method (\cite{Alek}) (see appendix) to find the regions describing instabilities. This method has been applied in the field of astrophysics very frequently (\cite{BnU06,BnU07a,Balbus2,Balbus1,UnR05,U03}) Instability exist if any of the conditions written below is satisfied.

\begin{equation}
\mu \Omega^{2}_{e} \omega^{3}_{B \Omega} < 0 \;,\;\;
\omega^{3}_{B \Omega} > 0 \;,\;\;
(\omega^{3}_{B \Omega})^{2}  > 0.
\end{equation}

From the above inequalities, it is clear that the instability will exist only when $\omega^{3}_{B \Omega} \neq 0$

It is to be noted that only those perturbations are considered here in which the wavevector is perpendicular to the magnetic field, i.e. $\vec{k} \cdot \vec{B}=0$.

Let us investigate the growth rate of the hydromagnetic instability. To calculate the growth rate of this instability,it is convenient to introduce the dimensionless quantities 
\begin{equation}\label{e:eq4}
\Gamma= \frac{\sigma}{\Omega_{e}} \;,\;\; \xi = 
\frac{1}{x^{2}} \frac{\omega^{2}_{0}}{\Omega^{2}_{e}} \;,\;\; 
\zeta = \frac{1}{x^{2}} \frac{\omega^{3}_{B \Omega}}{\Omega^{3}_{e}}
\;, \;\; x=ks
\end{equation}
The polynomial given by Eq. (\ref{e:eq4}) becomes 
\begin{equation}\label{e:eq5}
\Gamma^5 + \Gamma^3 (\xi x^2 + 1) + \Gamma^2 \zeta x^2 + \Gamma \mu \xi x^2 + \mu \zeta x^2 =0
\end{equation}

This equation is solved numerically (see \cite{Press} for details) by computing the eigen values of the matrix whose characteristic polynomial is given by equation (\ref{e:eq5})  for different values of  $\mu$, $\xi$ and $\zeta$ parameters.

Since we are interested in the instabilities, only real roots (perturbations $\propto \exp ( \sigma t - i \vec{k} \cdot \vec{r})$) are considered. We consider the parameter space constrained by $\zeta > \xi$ and investigate the growth of instability for different values of $\xi$ and $\zeta$, keeping in mind $\zeta>\xi$. It is to be noted that this parameter space was not considered by Bonanno and Urpin \cite{BnU06}. The Fig.(1), (2),\& (3) shows the dependence of Re ($\Gamma$) for $\mu = 0.3$. The value of $\zeta$ is fixed to $1.0$, $2.0$ and $3.0$ and the values of the parameter $\xi$ is varied for $\xi = 0.05,0.1,0.2$. Similar trend is observed in all the cases, i.e.  the growth rate of instability increases with the decrease in $\xi$.
For a value of $\xi = 0.05$, growth rate of the instability at $x^2=500$ increases from $3.5 \Omega_e$ to $4.6 \Omega_e$ and $5.5 \Omega_e$  for $\zeta=1.0,2.0,3.0$ respectively. As $x^2$ increases further, at $x^2=700$, the growth rate of instability increases from $3.8 \Omega_e$ for $\zeta=1.0$ to $5.1 \Omega_e$ for $\zeta=2$ to $6.0 \Omega_e$ for $\zeta=3$. It is evident that a vigorous instability is operating in the above described parameter space.

\begin{figure}
\includegraphics[width = 0.35\textwidth, angle=270]{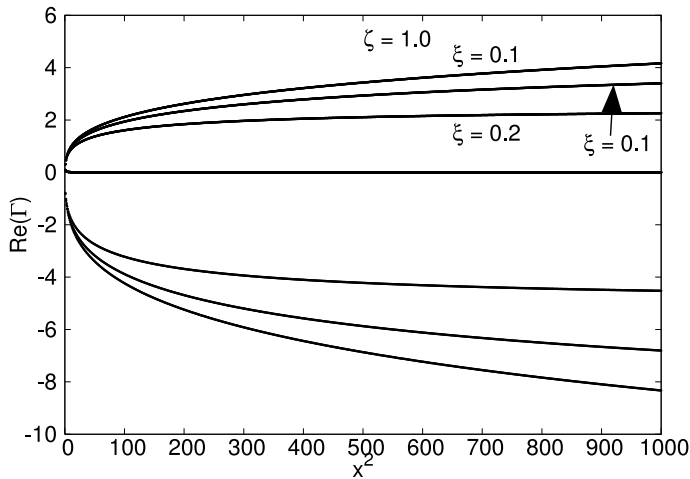}
\caption{The dependence of real parts of the root $\Gamma$ on $x^2$ for $\mu = 0.3$, $\zeta = 1.0$ and $\xi = 0.2,0.5,0.05$}
\end{figure}

\begin{figure}
\includegraphics[width = 0.35\textwidth, angle=270]{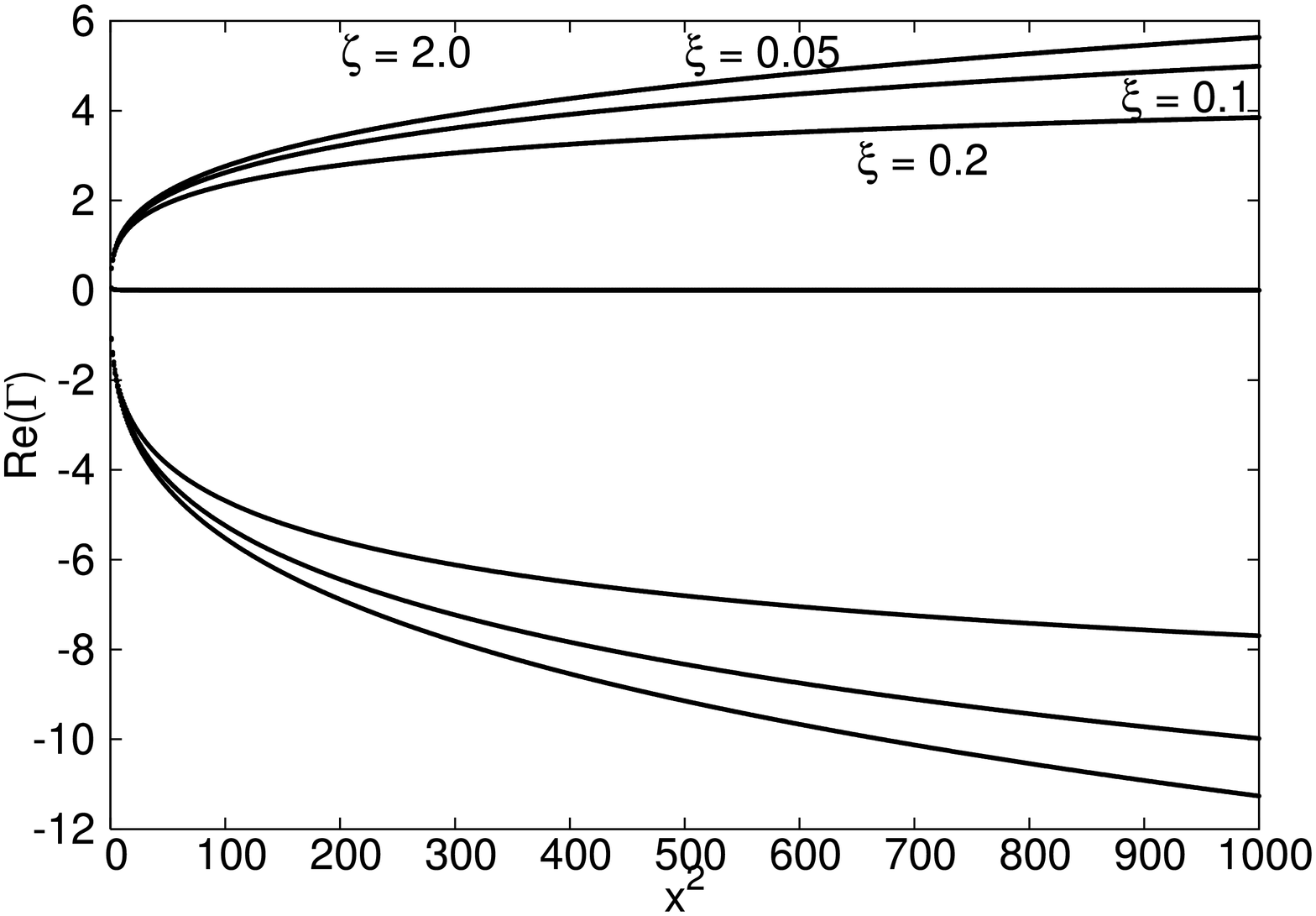}
\caption{The dependence of real parts of the root $\Gamma$ on $x^2$ for $\mu = 0.3$, $\zeta = 2.0$ and $\xi = 0.2,0.5,0.05$}
\end{figure}

\begin{figure}
\includegraphics[width = 0.35\textwidth, angle=270]{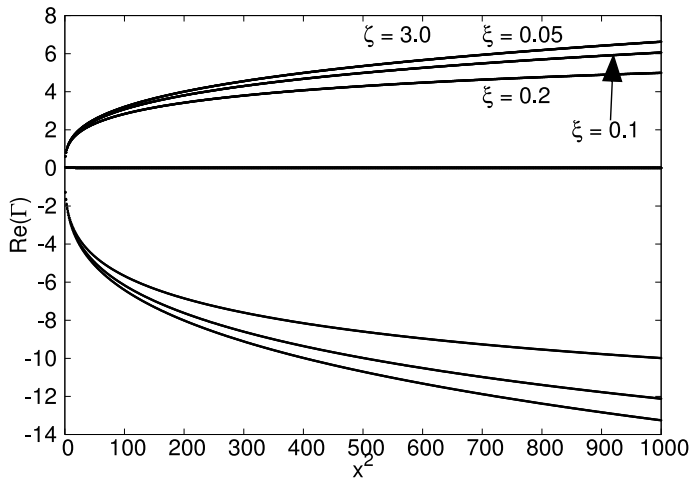}
\caption{The dependence of real parts of the root $\Gamma$ on $x^2$ for $\mu = 0.3$, $\zeta = 3.0$ and $\xi = 0.2,0.5,0.05$}
\end{figure}

\section{Results and Discussion} 
We revisit the problem of hydromagnetic instability in differentially rotating compressible flows and analyze the instability. The parameter space described by $\zeta>\xi$ was investigated and diskovered a vigorous instability. The behavior of the instability for a typical value of $\zeta$ and $\xi$ ( $\mu = 0.3, \zeta = 1.0,2.0,3.0, \xi = 0.05,0.1,0.2$) was studied. It is observed that the instability growth rate increases with the increase of $\zeta$ and it also depends on the wavenumber. The growth rate of the instability becomes very high, as high as ($\sim$ $6.0 \Omega$) for $x^2 =700$ at $\zeta =3$, $\xi = 0.05$).

In a recent work, Bonanno and Urpin \cite{BnU06} carried out the studies for a compressible flow and showed the presence of a new  instability. The notable point of their instability was that it existed for any value of magnetic field, the only condition being $\omega^{3}_{B \Omega} \neq 0$. The major difference between the studies carried out by Kim \& Ostriker \cite{kim}, Pessah \& Psaltis \cite {Pessah} and Bonanno \& Urpin \cite{BnU06} is that, in the later case, they considered a field with non vanishing radial and azimuthal component. 
We investigated a parameter space and noticed a high growth rate instability in that region. Recall that the MRI does not exist for $\vec{k} \cdot \vec{B}=0$. Since present study was carried out for $\vec{k} \cdot \vec{B}=0$, MRI does not operate in our case.  It should also be noted that for the incompressible case,i.e.  $c_s \rightarrow \infty$, polynomial (\ref{e:eq4}) becomes
\begin{equation}
\sigma \left (\sigma ^2 +\mu \Omega_e^2 \right)= 0
\end{equation}
It is clear that the present instability appears only for compressible fluid. It is to be noted that the radial component of magnetic field is non zero in the present case. Radial field  in the presence of differential rotation shears into an azimuthal field resulting in an increase in the magnitude of $B_{\phi}$. As an example, Desch \cite{Desch} describes that if $B_{\phi}$ = $B_r$ in a Keplerian disk at the beginning of one orbit, by the end of the orbit $B_{\phi}$ increases to $\sim 10 B_r$. After two orbits, $B_{\phi}$ increases even more, to $\sim 20 B_r$. This problem, i.e.  the increase in the azimuthal field in the present case was avoided by treating the state as a quasistationary state. When $\eta$ is small, one can obtain from Eq.~(4) that the azimuthal field grows  approximately linearly with time,
\begin{equation}
B_{\varphi}(t) = B_{\varphi}(0) + s \Omega' B_{s} t,
\end{equation} 
where $\Omega' = d \Omega/ds$, and $B_{\varphi}(0)$ is the azimuthal 
field at $t=0$. As long as the second term on the r.h.s. is small
compared to the first one, and
\begin{equation}
t \ll \tau_{\varphi} = \frac{1}{s \Omega'} \; 
\frac{B_{\varphi}(0)}{B_{s}},
\end{equation}
stretching of the azimuthal field does not affect significantly the basic state; $\tau_{\varphi}$ is the characteristic timescale of  generation of $B_{\varphi}$. As a result, the basic state can be treated as quasi-stationary during the time $t \ll \tau_{\varphi}$. In recent times, Bonanno and Urpin has shown that this instability exists in many cases of interest \cite{BnU07a,UB2}. The presence of this instability has also been demonstrated in the case of protostellar disks \cite{UB1} also. It is clear that the instability discussed in the present report appears in many astrophysical situations including protoplanatery disk (albeit with small growth rate).

\subsection{Difference between MRI and the present instability}
This instability is different and independent of the magnetorotational instability. The necessary and sufficient condition for this instability to occur is $\omega^{3}_{B \Omega} \neq 0$, i.e.  $B_{\varphi} B_{s} s \Omega' \neq 0, \Omega' = \frac{d \Omega}{ds}$; therefore, 

1) The instability exists for non vanishing radial and azimuthal component of magnetic field whereas the poloidal component of magnetic field is
important for  MRI to occur. MRI can exits for vanishing radial and azimuthal fields.

2) The instability exists for any differential rotation, i.e.  for any sign of $\frac{d\Omega}{ds}$ whereas MRI exists only when the angular velocity in the disk decreases outwards, i.e.  $\frac{d\Omega}{ds} < 0$.

3) The instability exists even in sufficiently strong magnetic field which suppresses the magnetorotational instability. MRI exists only in the weak field limit.

A special case was considered in the present report where wavevector is perpendicular to the magnetic field, i.e. $\vec{k} . \vec{B} = 0$. Since the growth rate of MRI is directly proportional to the  $\vec{k} . \vec{B}$, MRI does not exists in this case. This is one of the most important differences between the present instability and the MRI.  

\subsection{Where the instability operates}
It is observed that a high growth rate is reached for large values of $\zeta >\xi$ ($\zeta \geq 1$). If $s \frac {d\Omega}{ds} \sim \Omega $(typical in keplerian disk), the magnetic energy density is greater than the rotational energy density. Therefore,  the present instability will be operating in a regime where 
``magnetic energy density more that the  rotational energy density '' can be realized. For an accretion disk, the rotational energy (essentially by definition) is 
larger than all other energies  thermal, magnetic, radiation. So, the rotationally supported accretion disk are out of question to envisage the regions 
where magnetic energy density is more than rotational energy density. One possible realization of such regions might appear when one inches towards the inner most 
region of  accretion disk, where the disk starts truncating/evaporating and the rotational support starts getting diminishing and magnetic field takes over. A standard 
pulsar (or even more a magnetar) meets this condition. Consider a $10^{12}$ G surface field on a pulsar. Several scale heights away from the star the density is very 
small, so the rotational energy density is much less than the magnetic energy density. If we consider the low-density regions far above the disk then any magnetic 
bubble that floats away after reconnection could temporarily have a larger magnetic energy density than rotational energy density. 

The scenario closest to our regime was considered by Begelman and Pringle \cite {Begelman} where magnetic pressure exceeds the combined gas + radiation 
pressure in the disk but not the rotational energy density. For magnetohydrodynamic instabilities in accretion disk, the balance is achieved via the joint action 
of rotation, pressure gradients and magnetic tension. For the regions mentioned above, one can not  count on rotation. It is important to identify a robust 
equilibrium about which to perturb, perhaps gradients in pressure or radiation pressure could help in the  inner disk regions. One needs to  solve such scenario 
theoretically. We will consider this scenario in a future work.

\section{Conclusion}
In the present report, we have demonstrated the presence of very high growth rate instability in a differentially rotating compressible flows. It has been shown that the growth rate of this instability is very high and for a special case considered above, the growth rate increases to as high as $\sim 6 \Omega$ for $x^2=700$. It is observed that the present instability might be operating in regions where magnetic energy density is more than the rotational energy density.

\section*{Acknowledgments}
The author thanks C. Miller, V. Urpin and B.P.Pandey for the insightful discussions. The author also thanks both the anonymous referees for their useful comments. This research has made use of NASA's Astrophysics Data System and arXiv.org e-print archive. 

\appendix

\section{Hurwitz method}
\noindent Let us consider an $5^{th}$ order polynomial P(x)
\begin{equation}\label{e:app1}
P(x)= a_5 x^5 + a_4 x^4 + a_3 x^3 +a_2 x^2 +..... +a_0
\end{equation}
The Hurwitz theorem states that the above polynomial will be unstable if any of the following inequalities is satified \cite{UnR05}.
\begin{eqnarray}
\lefteqn{a_{0} <0 \;,}
\nonumber \\
\lefteqn{A_{1} \equiv a_{4} a_{3} - a_{2} < 0 \;, }
\nonumber \\
\lefteqn{A_{2} \equiv a_{2} (a_{4} a_{3} -a_{2}) - a_{4}(a_{4} a_{1} 
- a_{0}) <0 \;,}
\nonumber \\
\lefteqn{A_{3} \equiv (a_{4} a_{1} -a_{0}) [a_{2} (a_{4} a_{3} 
- a_{2}) - a_{4}
(a_{4} a_{1} - a_{0})] - }
\nonumber \\
&& \quad \quad \quad \quad \quad \quad - a_{0} (a_{4} a_{3} 
- a_{2})^{2} < 0 \;,
\label{13}
\end{eqnarray}

\label{lastpage}


\begin{thebibliography}{28}
\expandafter\ifx\csname natexlab\endcsname\relax\def\natexlab#1{#1}\fi
\expandafter\ifx\csname bibnamefont\endcsname\relax
  \def\bibnamefont#1{#1}\fi
\expandafter\ifx\csname bibfnamefont\endcsname\relax
  \def\bibfnamefont#1{#1}\fi
\expandafter\ifx\csname citenamefont\endcsname\relax
  \def\citenamefont#1{#1}\fi
\expandafter\ifx\csname url\endcsname\relax
  \def\url#1{\texttt{#1}}\fi
\expandafter\ifx\csname urlprefix\endcsname\relax\def\urlprefix{URL }\fi
\providecommand{\bibinfo}[2]{#2}
\providecommand{\eprint}[2][]{\url{#2}}

\bibitem[{\citenamefont{{Pringle}}(1981)}]{Pringle}
\bibinfo{author}{\bibfnamefont{J.~E.} \bibnamefont{{Pringle}}},
  \bibinfo{journal}{ARA\&A} \textbf{\bibinfo{volume}{19}}, \bibinfo{pages}{137}
  (\bibinfo{year}{1981}).

\bibitem[{\citenamefont{{Shakura} and {Syunyaev}}(1973)}]{Shakura}
\bibinfo{author}{\bibfnamefont{N.~I.} \bibnamefont{{Shakura}}}
  \bibnamefont{and} \bibinfo{author}{\bibfnamefont{R.~A.}
  \bibnamefont{{Syunyaev}}}, \bibinfo{journal}{Astron. Astrophys.}
  \textbf{\bibinfo{volume}{24}}, \bibinfo{pages}{337} (\bibinfo{year}{1973}).

\bibitem[{\citenamefont{{Heisenberg}}(1948)}]{Heisenberg}
\bibinfo{author}{\bibfnamefont{W.} \bibnamefont{{Heisenberg}}},
  \bibinfo{journal}{RSPSA} \textbf{\bibinfo{volume}{195}}, \bibinfo{pages}{402H}
  (\bibinfo{year}{1948}).

\bibitem[{\citenamefont{{Afshordi} et~al.}(2005)\citenamefont{{Afshordi},
  {Mukhopadhyay}, and {Narayan}}}]{Bani1}
\bibinfo{author}{\bibfnamefont{N.}~\bibnamefont{{Afshordi}}},
  \bibinfo{author}{\bibfnamefont{B.}~\bibnamefont{{Mukhopadhyay}}},
  \bibnamefont{and}
  \bibinfo{author}{\bibfnamefont{R.}~\bibnamefont{{Narayan}}},
  \bibinfo{journal}{ApJ} \textbf{\bibinfo{volume}{629}}, \bibinfo{pages}{373}
  (\bibinfo{year}{2005}), \eprint{arXiv:astro-ph/0412194}.

\bibitem[{\citenamefont{{Mukhopadhyay}
  et~al.}(2005)\citenamefont{{Mukhopadhyay}, {Afshordi}, and
  {Narayan}}}]{Bani2}
\bibinfo{author}{\bibfnamefont{B.}~\bibnamefont{{Mukhopadhyay}}},
  \bibinfo{author}{\bibfnamefont{N.}~\bibnamefont{{Afshordi}}},
  \bibnamefont{and}
  \bibinfo{author}{\bibfnamefont{R.}~\bibnamefont{{Narayan}}},
  \bibinfo{journal}{ApJ} \textbf{\bibinfo{volume}{629}}, \bibinfo{pages}{383}
  (\bibinfo{year}{2005}), \eprint{arXiv:astro-ph/0412193}.

\bibitem[{\citenamefont{{Honma}}(1996)}]{Honma}
\bibinfo{author}{\bibfnamefont{F.}~\bibnamefont{{Honma}}},
  \bibinfo{journal}{Pub. Astronom. Soc. Japan} \textbf{\bibinfo{volume}{48}},
  \bibinfo{pages}{77} (\bibinfo{year}{1996}).

\bibitem[{\citenamefont{{Manmoto} and {Kato}}(2000)}]{Manmoto}
\bibinfo{author}{\bibfnamefont{T.}~\bibnamefont{{Manmoto}}} \bibnamefont{and}
  \bibinfo{author}{\bibfnamefont{S.}~\bibnamefont{{Kato}}},
  \bibinfo{journal}{ApJ} \textbf{\bibinfo{volume}{538}}, \bibinfo{pages}{295}
  (\bibinfo{year}{2000}).

\bibitem[{\citenamefont{{Kato} and {Manmoto}}(2000)}]{Kato}
\bibinfo{author}{\bibfnamefont{S.}~\bibnamefont{{Kato}}} \bibnamefont{and}
  \bibinfo{author}{\bibfnamefont{T.}~\bibnamefont{{Manmoto}}},
  \bibinfo{journal}{ApJ} \textbf{\bibinfo{volume}{541}}, \bibinfo{pages}{889}
  (\bibinfo{year}{2000}).

\bibitem[{\citenamefont{{Gracia} et~al.}(2003)\citenamefont{{Gracia}, {Peitz},
  {Keller}, and {Camenzind}}}]{Gracia}
\bibinfo{author}{\bibfnamefont{J.}~\bibnamefont{{Gracia}}},
  \bibinfo{author}{\bibfnamefont{J.}~\bibnamefont{{Peitz}}},
  \bibinfo{author}{\bibfnamefont{C.}~\bibnamefont{{Keller}}}, \bibnamefont{and}
  \bibinfo{author}{\bibfnamefont{M.}~\bibnamefont{{Camenzind}}},
  \bibinfo{journal}{MNRAS} \textbf{\bibinfo{volume}{344}}, \bibinfo{pages}{468}
  (\bibinfo{year}{2003}), \eprint{arXiv:astro-ph/0301113}.

\bibitem[{\citenamefont{{Velikhov}}(1959)}]{Velikhov}
\bibinfo{author}{\bibfnamefont{E.~P.} \bibnamefont{{Velikhov}}},
  \bibinfo{journal}{Sov. Physics. JETP} \textbf{\bibinfo{volume}{9}},
  \bibinfo{pages}{995} (\bibinfo{year}{1959}).

\bibitem[{\citenamefont{{Chandrasekhar}}(1960)}]{Chandra}
\bibinfo{author}{\bibfnamefont{S.}~\bibnamefont{{Chandrasekhar}}},
  \bibinfo{journal}{Proceedings of the National Academy of Science}
  \textbf{\bibinfo{volume}{46}}, \bibinfo{pages}{253} (\bibinfo{year}{1960}).

\bibitem[{\citenamefont{{Balbus} and {Hawley}}(1991)}]{Balbus}
\bibinfo{author}{\bibfnamefont{S.~A.} \bibnamefont{{Balbus}}} \bibnamefont{and}
  \bibinfo{author}{\bibfnamefont{J.~F.} \bibnamefont{{Hawley}}},
  \bibinfo{journal}{ApJ} \textbf{\bibinfo{volume}{376}}, \bibinfo{pages}{214}
  (\bibinfo{year}{1991}).

\bibitem[{\citenamefont{{Hawley} et~al.}(1995)\citenamefont{{Hawley}, {Gammie},
  and {Balbus}}}]{Hawley1}
\bibinfo{author}{\bibfnamefont{J.~F.} \bibnamefont{{Hawley}}},
  \bibinfo{author}{\bibfnamefont{C.~F.} \bibnamefont{{Gammie}}},
  \bibnamefont{and} \bibinfo{author}{\bibfnamefont{S.~A.}
  \bibnamefont{{Balbus}}}, \bibinfo{journal}{ApJ}
  \textbf{\bibinfo{volume}{440}}, \bibinfo{pages}{742} (\bibinfo{year}{1995}).

\bibitem[{\citenamefont{{Sano} et~al.}(2004)\citenamefont{{Sano}, {Inutsuka},
  {Turner}, and {Stone}}}]{Hawley2}
\bibinfo{author}{\bibfnamefont{T.}~\bibnamefont{{Sano}}},
  \bibinfo{author}{\bibfnamefont{S.-i.} \bibnamefont{{Inutsuka}}},
  \bibinfo{author}{\bibfnamefont{N.~J.} \bibnamefont{{Turner}}},
  \bibnamefont{and} \bibinfo{author}{\bibfnamefont{J.~M.}
  \bibnamefont{{Stone}}}, \bibinfo{journal}{ApJ}
  \textbf{\bibinfo{volume}{605}}, \bibinfo{pages}{321} (\bibinfo{year}{2004}),
  \eprint{arXiv:astro-ph/0312480}.

\bibitem[{\citenamefont{{Blaes} and {Balbus}}(1994)}]{Blaes}
\bibinfo{author}{\bibfnamefont{O.~M.} \bibnamefont{{Blaes}}} \bibnamefont{and}
  \bibinfo{author}{\bibfnamefont{S.~A.} \bibnamefont{{Balbus}}},
  \bibinfo{journal}{ApJ} \textbf{\bibinfo{volume}{421}}, \bibinfo{pages}{163}
  (\bibinfo{year}{1994}).

\bibitem[{\citenamefont{{Kim} and {Ostriker}}(2000)}]{kim}
\bibinfo{author}{\bibfnamefont{W.-T.} \bibnamefont{{Kim}}} \bibnamefont{and}
  \bibinfo{author}{\bibfnamefont{E.~C.} \bibnamefont{{Ostriker}}},
  \bibinfo{journal}{ApJ} \textbf{\bibinfo{volume}{540}}, \bibinfo{pages}{372}
  (\bibinfo{year}{2000}), \eprint{arXiv:astro-ph/0004094}.

\bibitem[{\citenamefont{{Pessah} and {Psaltis}}(2005)}]{Pessah}
\bibinfo{author}{\bibfnamefont{M.~E.} \bibnamefont{{Pessah}}} \bibnamefont{and}
  \bibinfo{author}{\bibfnamefont{D.}~\bibnamefont{{Psaltis}}},
  \bibinfo{journal}{ApJ} \textbf{\bibinfo{volume}{628}}, \bibinfo{pages}{879}
  (\bibinfo{year}{2005}), \eprint{arXiv:astro-ph/0406071}.

\bibitem[{\citenamefont{{Bonanno} and {Urpin}}(2006)}]{BnU06}
\bibinfo{author}{\bibfnamefont{A.}~\bibnamefont{{Bonanno}}} \bibnamefont{and}
  \bibinfo{author}{\bibfnamefont{V.}~\bibnamefont{{Urpin}}},
  \bibinfo{journal}{PRE} \textbf{\bibinfo{volume}{73}}, \bibinfo{pages}{066301}
  (\bibinfo{year}{2006}).

\bibitem[{\citenamefont{{Sharma}}(2008)}]{Mradul}
\bibinfo{author}{\bibfnamefont{M.}~\bibnamefont{{Sharma}}},
  \bibinfo{journal}{MNRAS} \textbf{\bibinfo{volume}{391}},
  \bibinfo{pages}{1369} (\bibinfo{year}{2008}),
  \eprint{arxiv:astro-ph/0809.3125}.

\bibitem[{\citenamefont{{Aleksandrov} and {Laurentiev}}(1985)}]{Alek}
\bibinfo{author}{\bibfnamefont{A.}~\bibnamefont{{Aleksandrov}},
  \bibfnamefont{A.~{Kolmogorov}}} \bibnamefont{and}
  \bibinfo{author}{\bibfnamefont{M.}~\bibnamefont{{Laurentiev}}},
  \bibinfo{journal}{MIT Press, Cambridge}  (\bibinfo{year}{1985}).

\bibitem[{\citenamefont{{Bonanno} and {Urpin}}(2007{\natexlab{a}})}]{BnU07a}
\bibinfo{author}{\bibfnamefont{A.}~\bibnamefont{{Bonanno}}} \bibnamefont{and}
  \bibinfo{author}{\bibfnamefont{V.}~\bibnamefont{{Urpin}}},
  \bibinfo{journal}{ApJ} \textbf{\bibinfo{volume}{662}}, \bibinfo{pages}{851}
  (\bibinfo{year}{2007}{\natexlab{a}}).

\bibitem[{\citenamefont{{Balbus}}(2000)}]{Balbus2}
\bibinfo{author}{\bibfnamefont{S.~A.} \bibnamefont{{Balbus}}},
  \bibinfo{journal}{ApJ} \textbf{\bibinfo{volume}{534}}, \bibinfo{pages}{420}
  (\bibinfo{year}{2000}), \eprint{arXiv:astro-ph/9906315}.

\bibitem[{\citenamefont{{Balbus}}(2001)}]{Balbus1}
\bibinfo{author}{\bibfnamefont{S.~A.} \bibnamefont{{Balbus}}},
  \bibinfo{journal}{ApJ} \textbf{\bibinfo{volume}{562}}, \bibinfo{pages}{909}
  (\bibinfo{year}{2001}).

\bibitem[{\citenamefont{{Urpin} and {R{\"u}diger}}(2005)}]{UnR05}
\bibinfo{author}{\bibfnamefont{V.}~\bibnamefont{{Urpin}}} \bibnamefont{and}
  \bibinfo{author}{\bibfnamefont{G.}~\bibnamefont{{R{\"u}diger}}},
  \bibinfo{journal}{Astron. Astrophys} \textbf{\bibinfo{volume}{437}},
  \bibinfo{pages}{23} (\bibinfo{year}{2005}).

\bibitem[{\citenamefont{{Urpin}}(2003)}]{U03}
\bibinfo{author}{\bibfnamefont{V.}~\bibnamefont{{Urpin}}},
  \bibinfo{journal}{Astron. Astrophys} \textbf{\bibinfo{volume}{404}},
  \bibinfo{pages}{397} (\bibinfo{year}{2003}).

\bibitem[{\citenamefont{{Press} et~al.}(1992)\citenamefont{{Press},
  {Teukolsky}, {Vetterling}, and {Flannery}}}]{Press}
\bibinfo{author}{\bibfnamefont{W.~H.} \bibnamefont{{Press}}},
  \bibinfo{author}{\bibfnamefont{S.~A.} \bibnamefont{{Teukolsky}}},
  \bibinfo{author}{\bibfnamefont{W.~T.} \bibnamefont{{Vetterling}}},
  \bibnamefont{and} \bibinfo{author}{\bibfnamefont{B.~P.}
  \bibnamefont{{Flannery}}}, \emph{\bibinfo{title}{{Numerical recipes in
  FORTRAN. The art of scientific computing}}} (\bibinfo{publisher}{Cambridge:
  University Press, 2nd ed.}, \bibinfo{year}{1992}).

\bibitem[{\citenamefont{{Desch}}(2004)}]{Desch}
\bibinfo{author}{\bibfnamefont{S.~J.} \bibnamefont{{Desch}}},
  \bibinfo{journal}{ApJ} \textbf{\bibinfo{volume}{608}}, \bibinfo{pages}{509}
  (\bibinfo{year}{2004}).

\bibitem[{\citenamefont{{Bonanno} and {Urpin}}(2007{\natexlab{b}})}]{UB2}
\bibinfo{author}{\bibfnamefont{A.}~\bibnamefont{{Bonanno}}} \bibnamefont{and}
  \bibinfo{author}{\bibfnamefont{V.}~\bibnamefont{{Urpin}}},
  \bibinfo{journal}{PRE} \textbf{\bibinfo{volume}{76}}, \bibinfo{pages}{016303}
  (\bibinfo{year}{2007}{\natexlab{b}}), \eprint{arXiv:0706.2690}.

\bibitem[{\citenamefont{{Bonanno} and {Urpin}}(2008)}]{UB1}
\bibinfo{author}{\bibfnamefont{A.}~\bibnamefont{{Bonanno}}} \bibnamefont{and}
  \bibinfo{author}{\bibfnamefont{V.}~\bibnamefont{{Urpin}}},
  \bibinfo{journal}{A\&A} \textbf{\bibinfo{volume}{480}}, \bibinfo{pages}{27}
  (\bibinfo{year}{2008}), \eprint{arXiv:0801.1960}.

\bibitem[{\citenamefont{{Begelman} and {Pringle}}(2008)}]{Begelman}
\bibinfo{author}{\bibfnamefont{M.C.}~\bibnamefont{{Begelman}}} \bibnamefont{and}
  \bibinfo{author}{\bibfnamefont{J.E.}~\bibnamefont{{Pringle}}},
  \bibinfo{journal}{MNRAS} \textbf{\bibinfo{volume}{375}}, \bibinfo{pages}{1070}
  (\bibinfo{year}{2007}), \eprint{arXiv:0612300}.
\end{thebibliography}
\end{document}